\documentclass[prd,nofootinbib,preprintnumbers,floatfix]{revtex4}  
\usepackage[plainpages=false, colorlinks=true, anchorcolor=blue, linkcolor=blue, citecolor=blue, bookmarks=false]{hyperref}
\usepackage{url}
\usepackage{siunitx}
\newcommand{\rthis}[1]{\textcolor{black}{#1}}
\usepackage{graphicx}

\begin{document}
\author{Pragna  \surname{Mamidipaka}$^{1}$}%
 \altaffiliation{ee20btech11026@iith.ac.in}
 \author{Shantanu \surname{Desai}$^{2}$ }%
 \altaffiliation{shntn05@gmail.com}
\affiliation{$^{1}$ Department of Electrical Engineering, Indian Institute of Technology, Hyderabad, Kandi, Telangana-502284  India}

\title{Application of Efron-Petrosian method to radio pulsar fluxes}
\affiliation{$^{2}$ Department of Physics, Indian Institute of Technology, Hyderabad, Kandi, Telangana-502284  India}
\date{\today}

\begin{abstract}
 We apply the Efron-Petrosian technique to radio fluxes of pulsars detected in the Parkes multi-beam survey to test the independence of luminosity and distance. For this dataset, we find that 
 for four different distance exponents (ranging from 0.5 to 2), 
 the flux thresholds at which the luminosity and distances are uncorrelated,   correspond to very low $p$-values for the Kolmogorov-Smirnov test between the truncated and untruncated datasets. This is due to the fact that the  Parkes multi-beam survey  is not sufficiently homogeneous to lend itself to a treatment by the Efron-Petrosian method.
 We then repeat the analysis after rendering the dataset more homogeneous by excluding the distant pulsars from this sample. We find that for this culled dataset, the flux is consistent with distance exponents of 1.5 and 2.0.

\end{abstract}
\maketitle

\section{Introduction}

Pulsars are rotating neutron stars, which emit pulsed radio emissions with periods ranging from  milliseconds to a few seconds  with   magnetic fields ranging from $10^8$ to $10^{14}$ G~\cite{handbook,Reddy}. They  are known to be wonderful  laboratories for a  diverse suite of areas in Physics and Astronomy~\cite{Blandford92,KaspiKramer} from stellar evolution~\cite{Lorimer08}, dark matter~\cite{Kahya}, tests of equivalence principle~\cite{DesaiKahyaCrab} to Lorentz Invariance violation~\cite{DesaiLIV}.

In a series of recent works, Ardavan has applied the Efron-Petrosian (EP) technique~\cite{EP}
to probe the scaling dependence of the gamma ray fluxes  of pulsars~\cite{Ardavanfermi} (A22, hereafter)  as well as the  X-ray emission of magnetars (neutron stars with magnetic fields $> 10^{14}$ G)~\cite{Ardavanmagnetar}  as a function of distance. Their  analysis  of  three years of  Fermi-LAT data~\cite{Fermi2}  showed that a scaling of the  flux density ($F$)  with the distance ($D$) according to  $F \propto D^{-3/2}$,  is favored at higher levels of significance compared to the  inverse-square law scaling of $F \propto D^{-2}$. This was subsequently confirmed using the 12 year Fermi-LAT catalog~\cite{Fermi12yr} in ~\cite{Ardavan23}. A similar conclusion was also obtained~\cite{Ardavanmagnetar}  based on the analysis of X-ray fluxes of magnetars compiled in the McGill magnetar catalog~\cite{Kaspi14}.   These results also  agree with observational predictions of  the theoretical model for the pulsar emission mechanism outlined in \cite{Ardavan21}. If confirmed, this result is very exciting and would allow us to get insights into relativistic Plasma Physics in very strong magnetic fields and shed light on the pulsar emission mechanism, which is still not completely understood~\cite{Melrose}.

Previously,  there have been claims that the radio pulsar fluxes do not follow an inverse square law in models involving superluminal polarization
 currents~\cite{Ardavan09}. Such a model predicts an inverse linear scaling of the flux with distance. A claim was also made in literature  that the  radio pulsar fluxes {\it do}  scale inversely with the  first power of  distance according to ($F \propto D^{-1}$)~\cite{Singleton,Middleditch} based on the application of the Stepwise Maximum Likelihood Method~\cite{SWML}, in accord with the theoretical model proposed in ~\cite{Ardavan09}. However, this result could not be confirmed  using an independent analysis~\cite{Desai16}.

 Nevertheless, the original theoretical model proposed in ~\cite{Ardavan09} has been superseded by the model in ~\cite{Ardavan21}, which predicts that the fluxes of radio pulsars obey the inverse square law. Despite this, it behooves us to apply the same methodology as  in A22 to  radio pulsar fluxes.  The aim of this work is to  apply the EP method in the same way as A22 to  radio pulsars to confirm any putative violation of the inverse-square law, which has been argued for  gamma-ray fluxes. 

The outline of this manuscript is as follows. We recap the EP method and its myriad applications in literature in Sect.~\ref{sec:EP}. The dataset and analysis  for the observed radio pulsar population from the Parkes multibeam survey can be found in Sect.~\ref{sec:analysis}. 
We conclude in Sect.~\ref{sec:conclusions}. All logarithms in this manuscript are to the base 10.

\section{Efron-Petrosian method}
\label{sec:EP}
The EP method is a widely used technique in Astrophysics  and Cosmology  to account for selection biases or evolution  while dealing with  flux-limited or truncated samples~\cite{EP}. The EP method  has been applied to a diverse suite of astronomical objects such as quasars, gamma-ray bursts, magnetars, asteroids, solar flares, blazars etc~\cite{EP,LeePetrosian, Maloney,Wheatland,Kocevski,Dainotti13,Dainotti15,Dainotti22,Bargiacchi23}. This method has been used for a variety of science goals such as a probe of luminosity evolution, checking for intrinsic correlations between two astrophysical variables, precision tests of cosmological models, search for  cosmological time dilation, tests of standard candles, etc. More details on the myriad applications of the EP technique can be found in the aforementioned references. We provide a brief description of  the EP method. More details about this technique can be found in A22 or the above references.

Consider a flux limited catalog consisting of fluxes ($F$) obtained using a dedicated survey. If we assume that the pulsar flux scales with the distance ($D$) according to $F \propto D^{-\alpha}$, the isotropic luminosity ($L$) of the pulsar is given in terms of $F$ and $D$ according to~\cite{Ardavan23}:
\begin{equation}
L = 4 \pi l^2 \left(\frac{D}{l}\right)^{\alpha} F .
\label{eq:L}
\end{equation}
As discussed in A22, $l$ is a constant with dimensions of distance and is mainly used as a normalization constant \footnote{We note that in the pulsar literature instead of $L$ defined in Eq.~\ref{eq:L},  one usually defines a pseudo-luminosity, instead of  $L$ defined in Eq.~\ref{eq:L},  which does not contain the $4\pi$ factor~\cite{Bagchi} }. For the inverse-square law we get the familiar relation 
$L=4 \pi D^2 F$. 

If we now assume that a given pulsar survey has a flux threshold $S_{th}$, the corresponding truncation for the luminosity $L_{th}$ scales with $D$ according to
\begin{equation}
\log L_{th} = \log[ 4\pi l^{2-\alpha} S_{th}] + \alpha \log D
\end{equation}
For any distance-luminosity pair given by ($\log D_i$, $\log L_i$) one can find  a suitable  set  of luminosity-distance points given by: 
\begin{eqnarray}
  \log D &\leq& \log D_i~\rm{for}~i=1...n.  \\
  \log L &\geq& \log [4 \pi  l ^{2-\alpha} S_{th}] +   \alpha \log D ,
\end{eqnarray}
where $n$ is the number of pulsars not excluded by the flux threshold.
All ($D,L$) pairs which satisfy the above conditions are referred to  as  ``associated''~\cite{Dainotti22} or ``comparable''~\cite{Ardavanfermi}  set to ($D_i,L_i$). The total number of such associated pairs corresponding to ($D_i,L_i$) is equal to $N_i$. We now determine the $y$-rank  ($L_i$) of this point  ($\mathcal{R}_i$) compared to  its associated set of points, when  ranked  according to ascending order.

The EP technique then computes the following normalized statistic (which is related to the  Kendall-$\tau$ statistic~\cite{Dainotti13}) for all data points ($n$) greater than the flux threshold:
\begin{equation}
\tau = {{\sum \limits_{i=1}^n{(\mathcal{R}_i-\mathcal{E}_i)}} \over {\sqrt{\sum \limits_{i=1}^n{\mathcal{V}_i}}}}, 
\end{equation}
where $\mathcal{E}_i=\frac{1}{2}(N_i+1)$ and  $\mathcal{V}_i=\frac{1}{12}(N_i^2-1)$ 
The hypothesis of independence between $L$ and $D$ depends on the absolute value of $\tau$.
If $L$ and $D$ are independent, then the value of $\tau$ should be close to 0. If they are correlated the values of $\tau$ are quite high and the hypothesis of independence can be rejected at high significance levels.   One can quantify this hypothesis of rejection of independence of distance and luminosity by  calculating a  $p$-value, given by~\cite{EP}:
\begin{equation}
p = \left(\frac{2}{\pi}\right)^{1/2} \int_{|\tau|}^{\infty} \exp(-x^2/2) dx
\label{eq:pvalue}
\end{equation}
In terms of $Z$-score, one can reject the hypothesis that $L$ and $D$ are independent at significance equal to $n\sigma$  if $|\tau| > n$. In the literature, the EP method has been used in a couple of different ways. One way is to scale the luminosity with some power law of distance (or redshift) and find the distance  exponent  for which $\tau=0$ (or $|\tau|<1$~\cite{Dainotti13}), which corresponds to the independence of the corrected luminosity and distance~\cite{Bargiacchi23}. Alternately, one can compare the relative significance for the independence of hypothesis for different distance exponents. This is the approach adopted in A22 (and other recent works by Ardavan). We also note that the original EP method does not have any specific recommendations about what value of flux threshold to use. The original EP method recommended to use  the instrumental limiting sensitivity for the flux threshold~\cite{Maloney}. However, since sometimes this  maybe too low,  an elevated value for  the flux threshold has often been used~\cite{Dainotti13}.
Whether any one of these thresholds are relevant or not is dictated partly by the physics that underlies the problem, partly by the value of the probability of detection that these thresholds imply~\cite{Bryant} and partly by the requirement that the cut dataset and the uncut dataset should both be drawn from the same parent distribution (the unknown distribution that is complete over all values of the flux). 
It has been pointed out  that one gets physically meaningful results for the EP method when the detection thresholds are chosen near the peak of the dataset histogram~\cite{Bryant}. To determine the range of values of the detection threshold for which this latter requirement is satisfied,  one can apply the Kolmogorov-Smirnov (KS)  test to the dataset under consideration~\cite{Bryant,DPB21,Ardavanfermi,Ardavanmagnetar,Ardavan23}. ~\citet{Bryant} have  shown many examples from literature where  misleading results were obtained, because the flux thresholds are not chosen correctly. In most cases,  one  needs to   evaluate $\tau$ as a function of  flux threshold  over a wide range of thresholds and check for the corresponding KS test $p$-value at these flux thresholds in addition to the value of $\tau$. This is the approach followed in the recent works by Ardavan.

\section{Dataset and analysis for  observed  radio pulsar fluxes}
\label{sec:analysis}

At the time of writing there are a total of 3389 pulsars in the v1.70 ATNF pulsar catalog~\cite{ATNF}. Some of these pulsars (for example the first discovered radio pulsar) have been discovered serendipitiously, while there  is a lot of heterogeneity in the surveys used to detect the pulsars.
Therefore, to avoid any systematics related to different  flux thresholds which could lead to biased results~\cite{Bryant}, instead of applying the EP method to all pulsars, we decided to apply it to pulsars from only one specific survey, viz. the Parkes multi-beam survey~\cite{PMBS}. The Parkes multi-beam survey carried out a survey of the galactic plane with $|b| < 5^{\circ}$ and $l=260^{\circ}$ to $l=50^{\circ}$. The observations were made using a 13-beam receiver on the 64 m Parkes radio telescope with two polarizations per beam  at a frequency of 1374 MHz covering a bandwidth of 288 MHz. It has a limiting flux sensitivity of 0.2 mJy.

We first downloaded all the pulsars tagged as {\tt pksmb} (detected in the Parkes multi-beam survey) in the ATNF catalog. In this way, we obtain a  total of 1124 pulsars. Considering only the pulsars with period $ > 0.01s$ (to remove the millisecond pulsars), we get 1095 pulsars. Further, after removing pulsars without valid flux and distance estimates, we get 1071 pulsars. For each pulsar, we downloaded  the pulsar distances and the flux measured at 1400 MHz ($S_{1400}$), which was then  converted  from mJy to ergs $\rm{cm^{-2} s^{-1}}$. In the ATNF catalog, the distances have been obtained from the dispersion measure using the YMW16 electron density model~\cite{YMW16}.

\begin{figure}[hbt!]
     \centering
        \includegraphics[width=0.9\textwidth]{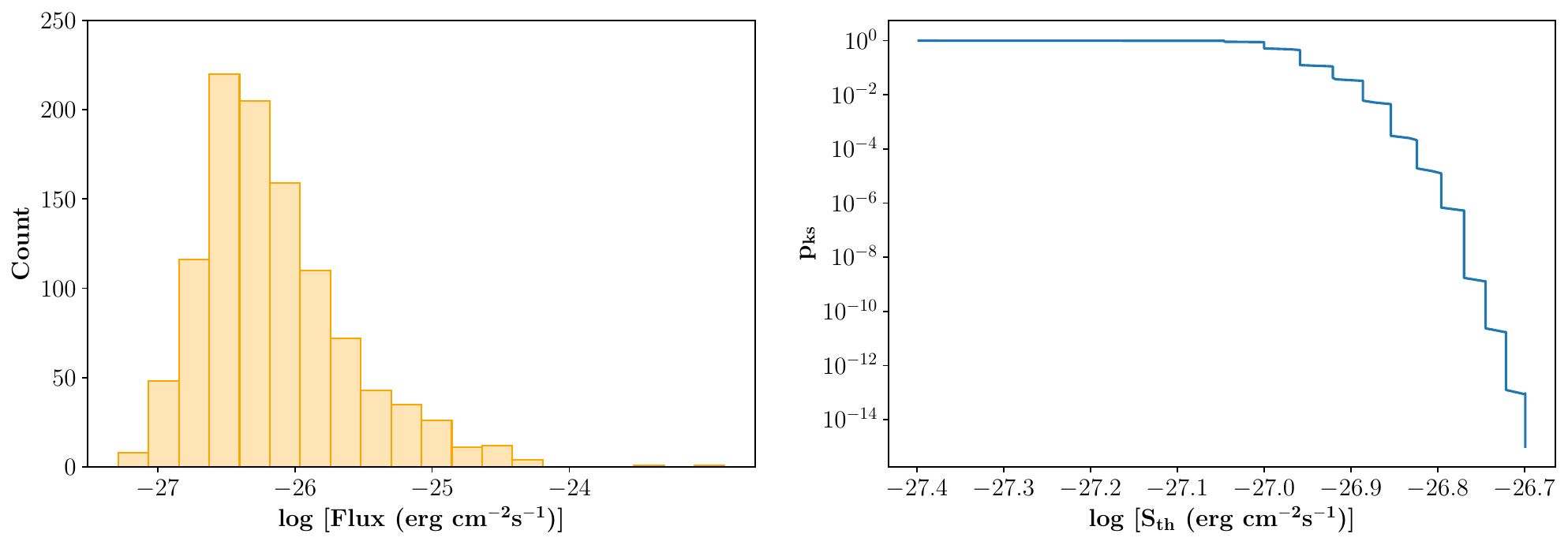}
            \caption{\label{fig:1a} (a) Histogram of the logarithm of radio fluxes of pulsars discovered in the Parkes multi-beam survey  at 1400 MHz.  (b)  $p$-value from the KS test ($p_{KS}$) as a function of flux threshold, for testing whether the truncated versions of the dataset (after discarding pulsars according to a flux threshold) are drawn from the same distribution.} 
\end{figure}

\begin{figure}[hbt!]
     \centering
        \includegraphics[width=0.625\textwidth]{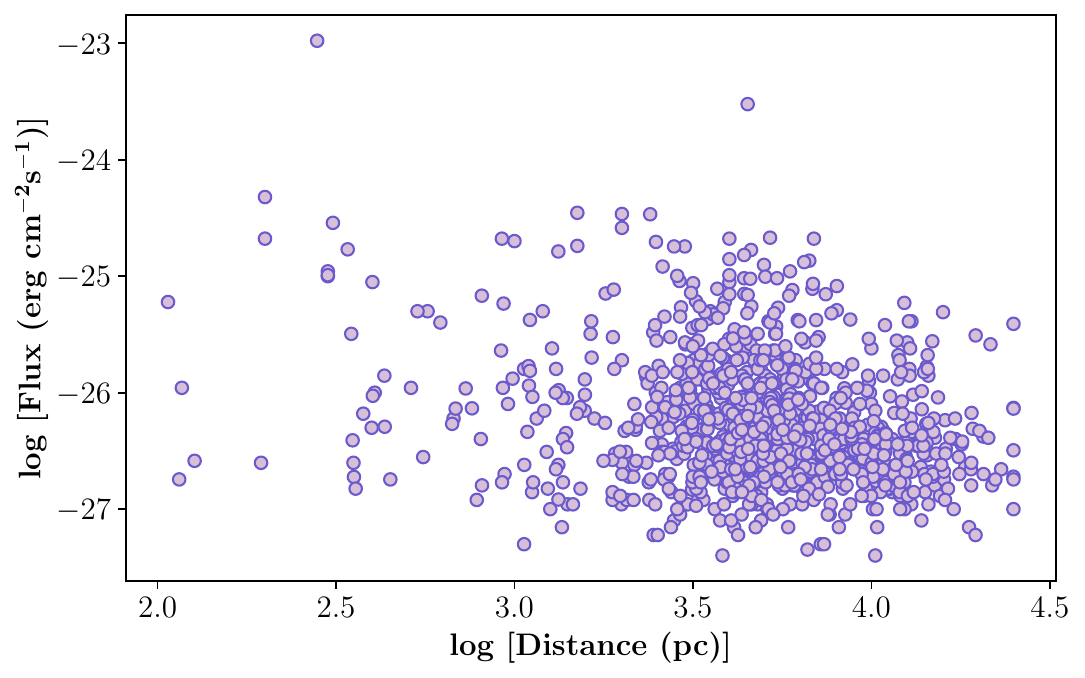}
            \caption{Scatter plot of the logarithm of radio pulsar flux as a function of the  logarithm of distance.}
            \label{fig:1b}
\end{figure}

\begin{figure}[hbt!]
     \centering
        \includegraphics[width=0.56\textwidth]{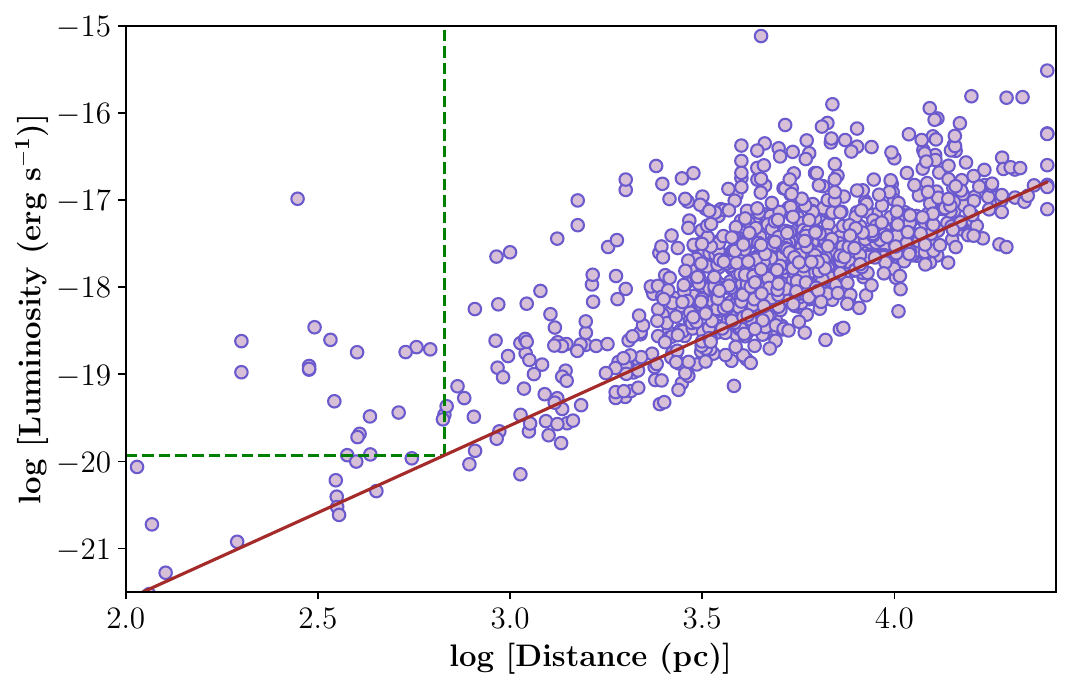}
            \caption{The corresponding Luminosity-distance dataset for the Flux-Distance dataset in Fig~\ref{fig:1b}. The solid red line represents the threshold by $\log (S_{th})=-26.690$. Those elements of this data set that lie within (and on the boundary of) the rectangular area bounded by the vertical axis and the vertical and horizontal dashed lines (in green) comprise the set comparable to the data point (2.82, -19.46) on the vertical dashed line. }
        \label{fig:2a}    
            \end{figure}

The distribution of  the logarithm of $S_{1400}$ can be found in Fig.~\ref{fig:1a}. 
Corresponding to each of these flux thresholds, we use  the two-sample KS test~\cite{astroml} between  the truncated dataset, consisting of pulsars having flux greater than the given  threshold  and the original untruncated dataset. The probability that both datasets are drawn from the same distribution can be characterized by the $p$-value from the KS test\footnote{We computed the two-sample KS test $p$-value using {\tt scipy} in {\tt Python2}.}, which we refer to as $p_{KS}$.
This plot of $p_{KS}$ as a function of the flux threshold  can be found in the right panel of Fig.~\ref{fig:1a}. We find that the KS $p$-value decreases with increasing threshold, but turns around for flux thresholds above $\log(S_{th})>-26$ and asymptotes to values of around $0.1$ at the maximum observed fluxes. We note that  the peak of the flux histogram ($\log (S)\sim -26.4$) corresponds to a very low value of  $p_{KS} =4.75 \times 10^{-71}$.
The distribution of the logarithm of flux as a function of the distance can be found in Fig.~\ref{fig:1b}. The distribution of the luminosity (assuming an inverse square law) can be found in Fig.~\ref{fig:2a}. The solid red line in Fig.~\ref{fig:2a} shows the flux threshold marked for  a flux value of $2.04 \times 10^{-27} \rm{erg  cm^{-2}s^{-1}}$. The points below the red line are excluded from the EP analysis. The rectangular area indicated by the dashed lines   shows the set comparable to one fiducial value of the distance and luminosity.

\begin{figure}[hbt!]
     \centering
        \includegraphics[width=0.6\textwidth]{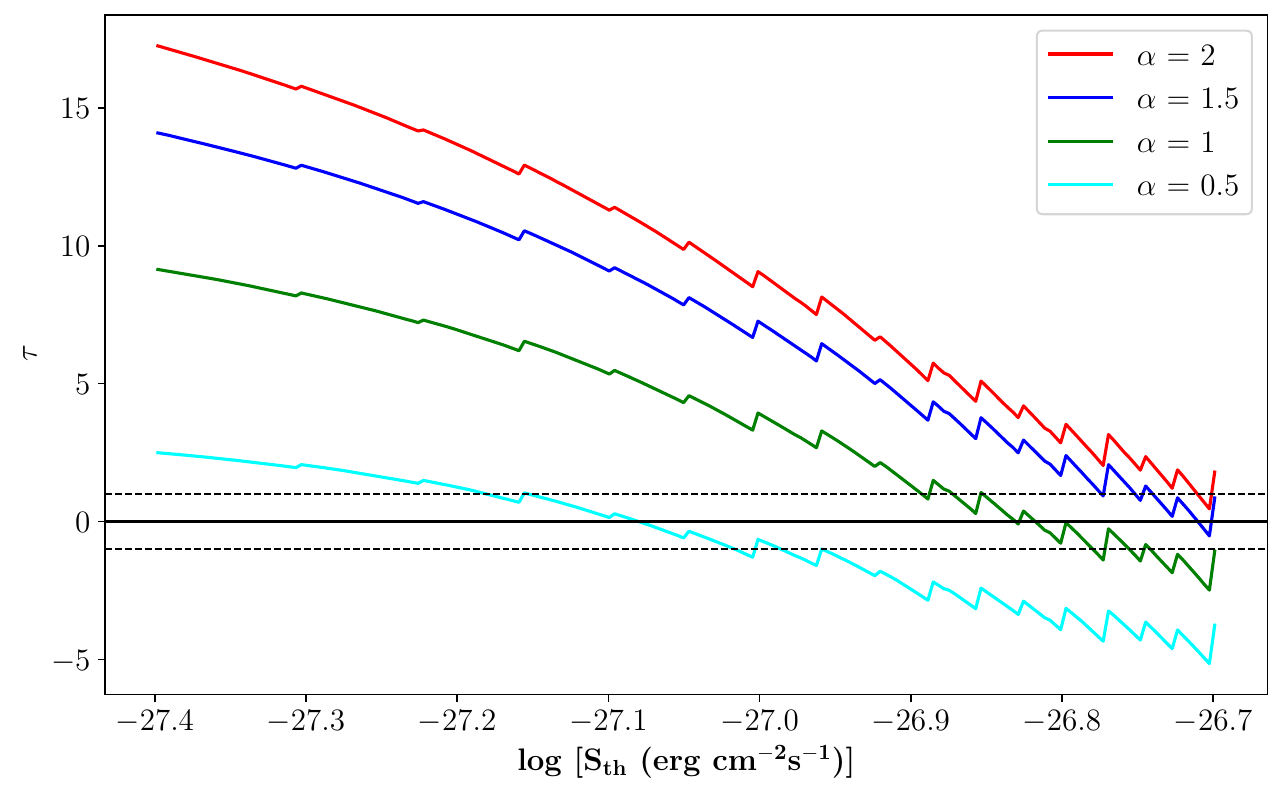}
            \caption{The Efron–Petrosian statistic $\tau$ versus the logarithm of the flux threshold, for different values of $\alpha$ for Parkes multibeam survey pulsars. The dashed lines correspond to  values of $\tau = \pm 1$. \rthis{For $\alpha = 2$, the flux threshold touches zero at $-26.56$, which lies outside the range of this plot.} }
            \label{fig:thres}
\end{figure}

\begin{figure}[hbt!]
     \centering
        \includegraphics[width=0.9\textwidth]{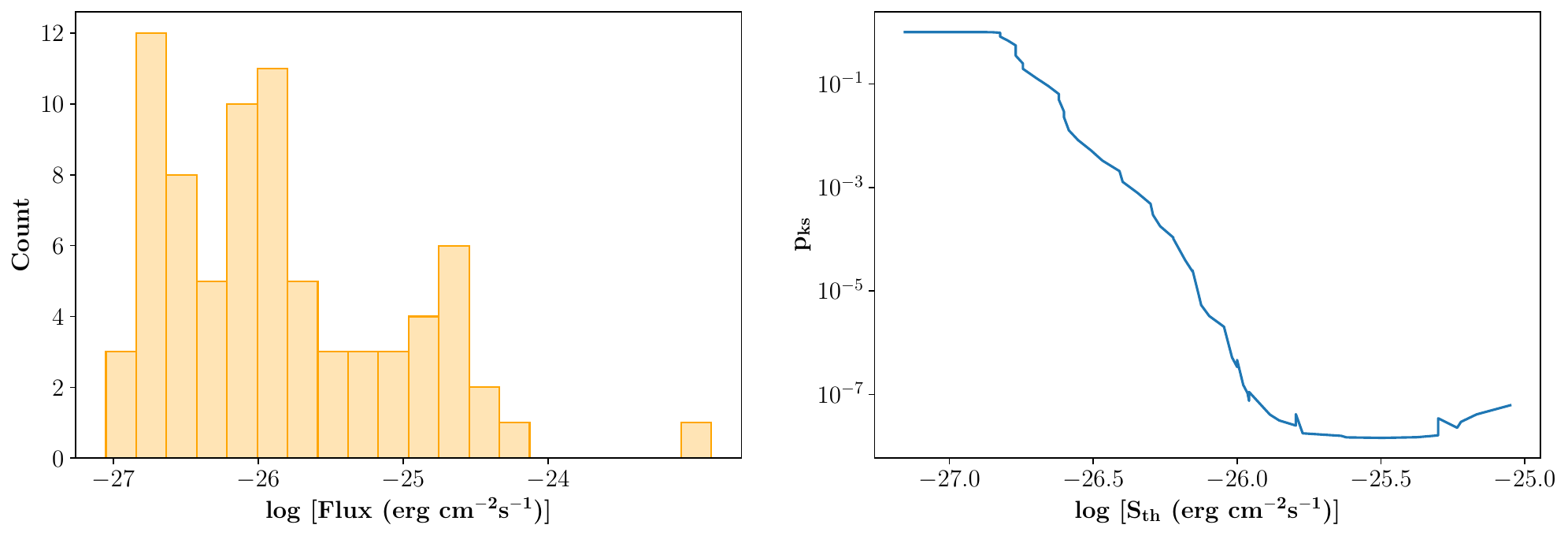}
            \caption{\label{fig:5} (a) Histogram of the logarithm of radio fluxes of pulsars, for the reduced dataset with $\log(D) < 3.2 $.  (b)  $p$-value from the KS test ($p_{KS}$) as a function of flux threshold, for testing whether the truncated versions of the dataset (after discarding pulsars according to a flux threshold) are drawn from the same distribution.} 
\end{figure}

\begin{figure}[hbt!]
     \centering
        \includegraphics[width=0.6\textwidth]{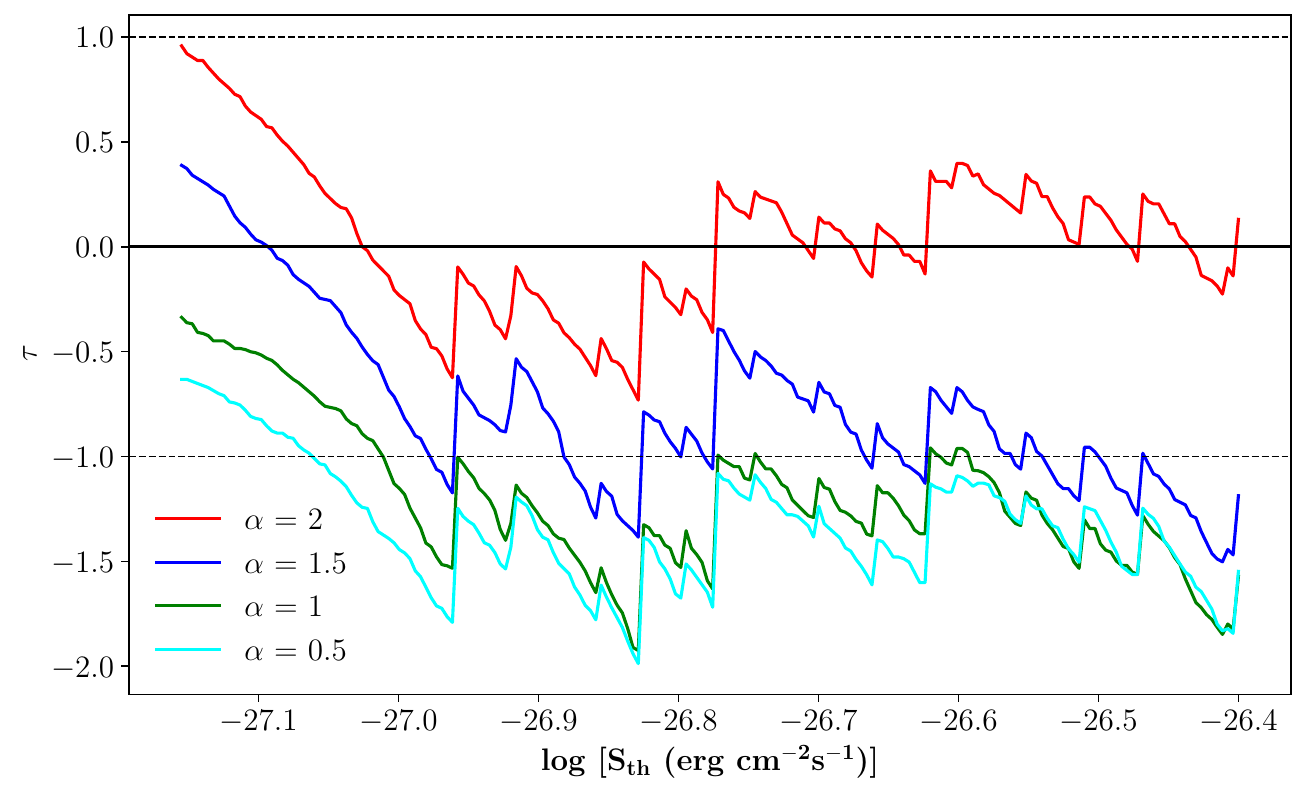}
            \caption{The Efron–Petrosian statistic $\tau$ versus the logarithm of the flux threshold for the culled sample of Parkes multibeam survey pulsars (with $\log(D) < 3.2 $), for different values of $\alpha$. The dashed lines correspond to  values of $\tau = \pm 1$. For $\alpha=2$, $\tau=0$ at $\log (S_{th})  = -27.02$ corresponding to  $p_{KS}$ at equal to  $1$. For $\alpha=1.5$, $\tau=0$ at $\log (S_{th}) = -27.09$, with $p_{KS}$  equal to  $1$. For $\alpha=1$ and 0.5, $\tau=0$ at  $\log (S_{th})= -25.85$ ($\alpha=1$) and $\log (S_{th})= -25.79$ ($\alpha=0.5$)   which corresponds to   $p_{KS}=3.15 \times 10^{-8}$ and $2.51 \times 10^{-8}$, respectively. Therefore, for this culled dataset, the pulsar fluxes  are   consistent with the  distance exponents of 1.5 and 2.}
            \label{fig:thres_culled}
\end{figure}

The distribution of the $\tau$ value for different values of $\alpha$ as a function of the flux threshold can be found in Fig.~\ref{fig:thres}. 
The values of the flux threshold for which $\tau$ becomes zero for each of the exponents are as follows. For $\alpha=2$, $\tau=0$ at $\log (S_{th})  = -26.56$. The $p$-value of KS test at this flux threshold is equal to  $1.06 \times 10^{-39}$. For $\alpha=1.5$, $\tau=0$ at $\log (S_{th}) = -26.72$. The corresponding $p$-value is equal to  $1.68 \times 10^{-11}$. For $\alpha=1$ and $\alpha=0.5$, $\tau=0$ at  $\log (S_{th})= -26.83$ ($\alpha=1$) and $\log (S_{th})= -27.082$ ($\alpha=0.5$),   which correspond $p_{ks}$ of  $2.11 \times 10^{-4}$ and  $9.86 \times 10^{-1}$, respectively. Hence, the KS $p$-values are very small for the  values of the flux at  which $\tau$ is equal to zero for all distance exponents, except $\alpha = 0.5$.

Therefore, prima-facie, we cannot draw any conclusions about  the correct distance exponent, since we get very small $p-$values from the KS test, for the fluxes at which the luminosity and distances are uncorrelated  for the higher distance exponents. We get a reasonable value of $P_{ks}$ for an unacceptably low value of $\alpha$, i.e., for $\alpha = 0.5$. However, this does not point to a failing or deficiency of the method. From Fig.~\ref{fig:1a} we find that there is  a clustering of pulsars for $\log (D (pc))>$ 3.2.
Therefore, the pulsars in the Parkes multi-beam survey data are not distributed uniformly with distance, because of which the dataset to which we have applied the EP method is not sufficiently homogeneous.

\subsection{Analysis with a culled dataset with $\log (D) < 3.2$ }
In order to apply EP analysis, we created a subsample of the aforementioned Parkes multi-beam survey with $\log (D) < 3.2$, where $D$ is expressed in pc,  so that the dataset is nearly homogeneous as a function of distance.  The flux histogram of this culled sample along with $p_{KS}$ as a function of the flux threshold  can be found in Fig.~\ref{fig:5}. We then redid the same EP analysis as what was done for the full sample. This plot of EP $\tau$ as a function of the flux threshold can be found in Fig.~\ref{fig:thres_culled}. Once again, we checked the values of $p_{KS}$ at which $\tau=0$ for all the four distance exponents. \rthis{For $\alpha=2$, $\tau=0$ for $\log (S_{th})  = -27.02$ with $p_{KS}=1$. For $\alpha=1.5$, $\tau=0$ at $\log (S_{th}) = -27.09$, with $p_{KS}=1$. For $\alpha=1$, $\tau=0$ at  $\log (S_{th})= -25.85$, with $p_{KS}=3.15 \times 10^{-8}$. Finally for $\alpha=0.5$, $\tau=0$ at $\log (S_{th})= -25.79$, with $p_{KS}=2.51\times 10^{-8}$.} Therefore, the  values of the flux threshold
at which $\tau=0$ for  distance exponents of 0.5 and 1 correspond to unphysically low values of $p_{KS}$, implying that these distance exponents are not viable. However, the corresponding values for the distance exponents of 1.5 and 2.0 correspond to $p_{KS}$ of one, implying the truncated and the original dataset are similar. This shows the fluxes of pulsars for the culled Parkes multi-beam survey dataset with $\log (D) < 3.2$ is consistent  with distance exponents of 1.5 and 2.0.

\section{Conclusions}
\label{sec:conclusions}
Recently, in a series of works Ardavan has demonstrated  that the gamma-ray flux of pulsars as well as the X-ray flux of magnetars show a violation of the inverse-square law and the flux scales with distance according to $F \propto D^{-3/2}$. This conclusion was based on the application of  the  EP method, where it was found that the independence of luminosity and distance is rejected at a higher level of significance for an inverse-square law scaling compared to $D^{-3/2}$. These results agree with the theoretical predictions in ~\cite{Ardavan21} for the X-ray and gamma-ray fluxes. However, the same  theoretical model predicts an inverse-square law  for the radio fluxes.

We then replicated this procedure on the radio pulsar fluxes at 1400 MHz, for pulsars discovered with the Parkes multi-beam survey. 
We found that the flux values for which the pulsar luminosity and distances are uncorrelated (using four different distance exponents)
correspond to very low $p$-values for the KS test between the truncated and untruncated pulsar sample. This is due to the fact that Parkes multi-beam survey dataset  is not sufficiently homogeneous to lend itself to a treatment by the EP method, because of clustering of pulsars at large distances. 

Therefore, we rendered the data more homogeneous by creating a subset of the above dataset  by removing the distant pulsars, and  only using those pulsars with  $\log(D)< 3.2$ and repeated the same analysis. We find that once again the values of the flux threshold for which  the distance and luminosity are uncorrelated correspond to unphysically low $p$-values for the KS test for distance exponents of 0.5 and 1. However, the corresponding flux thresholds for distance exponents of 1.5 and 2 correspond to $p$-values of one. This shows that for this culled subsample of Parkes multi-beam survey data obtained after excluding distant pulsars, the flux is consistent with both the inverse-square law as well as with a flux scaling of $F \propto D^{-1.5}$.

In the spirit of open science, we have made our analysis codes publicly available and these codes can be found at \url{https://github.com/Pymamid/EP-statistic-radio-pulsars.git}.

\begin{acknowledgements}
We are grateful to Maria Giovanna Dainotti and Manjari Bagchi for useful correspondence, as well as  the anonymous  referee for several useful comments and constructive feedback on our manuscript.
\end{acknowledgements}

\bibliography{main}

\begin{thebibliography}{37}
\expandafter\ifx\csname natexlab\endcsname\relax\def\natexlab#1{#1}\fi
\expandafter\ifx\csname bibnamefont\endcsname\relax
  \def\bibnamefont#1{#1}\fi
\expandafter\ifx\csname bibfnamefont\endcsname\relax
  \def\bibfnamefont#1{#1}\fi
\expandafter\ifx\csname citenamefont\endcsname\relax
  \def\citenamefont#1{#1}\fi
\expandafter\ifx\csname url\endcsname\relax
  \def\url#1{\texttt{#1}}\fi
\expandafter\ifx\csname urlprefix\endcsname\relax\def\urlprefix{URL }\fi
\providecommand{\bibinfo}[2]{#2}
\providecommand{\eprint}[2][]{\url{#2}}

\bibitem[{\citenamefont{{Lorimer} and {Kramer}}(2012)}]{handbook}
\bibinfo{author}{\bibfnamefont{D.~R.} \bibnamefont{{Lorimer}}}
  \bibnamefont{and} \bibinfo{author}{\bibfnamefont{M.}~\bibnamefont{{Kramer}}},
  \emph{\bibinfo{title}{{Handbook of Pulsar Astronomy}}}
  (\bibinfo{year}{2012}).

\bibitem[{\citenamefont{{Reddy Ch.} and {Desai}}(2022)}]{Reddy}
\bibinfo{author}{\bibfnamefont{T.~T.} \bibnamefont{{Reddy Ch.}}}
  \bibnamefont{and} \bibinfo{author}{\bibfnamefont{S.}~\bibnamefont{{Desai}}},
  \bibinfo{journal}{New Astronomy} \textbf{\bibinfo{volume}{91}},
  \bibinfo{eid}{101673} (\bibinfo{year}{2022}), \eprint{2011.03771}.

\bibitem[{\citenamefont{{Blandford}}(1992)}]{Blandford92}
\bibinfo{author}{\bibfnamefont{R.~D.} \bibnamefont{{Blandford}}},
  \bibinfo{journal}{Philosophical Transactions of the Royal Society of London
  Series A} \textbf{\bibinfo{volume}{341}}, \bibinfo{pages}{177}
  (\bibinfo{year}{1992}).

\bibitem[{\citenamefont{{Kaspi} and {Kramer}}(2016)}]{KaspiKramer}
\bibinfo{author}{\bibfnamefont{V.~M.} \bibnamefont{{Kaspi}}} \bibnamefont{and}
  \bibinfo{author}{\bibfnamefont{M.}~\bibnamefont{{Kramer}}},
  \bibinfo{journal}{arXiv e-prints} \bibinfo{eid}{arXiv:1602.07738}
  (\bibinfo{year}{2016}), \eprint{1602.07738}.

\bibitem[{\citenamefont{{Lorimer}}(2008)}]{Lorimer08}
\bibinfo{author}{\bibfnamefont{D.~R.} \bibnamefont{{Lorimer}}},
  \bibinfo{journal}{Living Reviews in Relativity}
  \textbf{\bibinfo{volume}{11}}, \bibinfo{eid}{8} (\bibinfo{year}{2008}),
  \eprint{0811.0762}.

\bibitem[{\citenamefont{{Desai} and {Kahya}}(2016)}]{Kahya}
\bibinfo{author}{\bibfnamefont{S.}~\bibnamefont{{Desai}}} \bibnamefont{and}
  \bibinfo{author}{\bibfnamefont{E.~O.} \bibnamefont{{Kahya}}},
  \bibinfo{journal}{Modern Physics Letters A} \textbf{\bibinfo{volume}{31}},
  \bibinfo{eid}{1650083} (\bibinfo{year}{2016}), \eprint{1510.08228}.

\bibitem[{\citenamefont{{Desai} and {Kahya}}(2018)}]{DesaiKahyaCrab}
\bibinfo{author}{\bibfnamefont{S.}~\bibnamefont{{Desai}}} \bibnamefont{and}
  \bibinfo{author}{\bibfnamefont{E.}~\bibnamefont{{Kahya}}},
  \bibinfo{journal}{European Physical Journal C} \textbf{\bibinfo{volume}{78}},
  \bibinfo{eid}{86} (\bibinfo{year}{2018}), \eprint{1612.02532}.

\bibitem[{\citenamefont{{Desai}}(2023)}]{DesaiLIV}
\bibinfo{author}{\bibfnamefont{S.}~\bibnamefont{{Desai}}},
  \bibinfo{journal}{arXiv e-prints} \bibinfo{eid}{arXiv:2303.10643}
  (\bibinfo{year}{2023}), \eprint{2303.10643}.

\bibitem[{\citenamefont{{Efron} and {Petrosian}}(1992)}]{EP}
\bibinfo{author}{\bibfnamefont{B.}~\bibnamefont{{Efron}}} \bibnamefont{and}
  \bibinfo{author}{\bibfnamefont{V.}~\bibnamefont{{Petrosian}}},
  \bibinfo{journal}{\apj} \textbf{\bibinfo{volume}{399}}, \bibinfo{pages}{345}
  (\bibinfo{year}{1992}).

\bibitem[{\citenamefont{{Ardavan}}(2022{\natexlab{a}})}]{Ardavanfermi}
\bibinfo{author}{\bibfnamefont{H.}~\bibnamefont{{Ardavan}}},
  \bibinfo{journal}{arXiv e-prints} \bibinfo{eid}{arXiv:2201.09256}
  (\bibinfo{year}{2022}{\natexlab{a}}), \eprint{2201.09256}.

\bibitem[{\citenamefont{{Ardavan}}(2022{\natexlab{b}})}]{Ardavanmagnetar}
\bibinfo{author}{\bibfnamefont{H.}~\bibnamefont{{Ardavan}}},
  \bibinfo{journal}{arXiv e-prints} \bibinfo{eid}{arXiv:2202.05162}
  (\bibinfo{year}{2022}{\natexlab{b}}), \eprint{2202.05162}.

\bibitem[{\citenamefont{{Abdo} et~al.}(2013)\citenamefont{{Abdo}, {Ajello},
  {Allafort}, {Baldini}, {Ballet}, {Barbiellini}, {Baring}, {Bastieri},
  {Belfiore}, {Bellazzini} et~al.}}]{Fermi2}
\bibinfo{author}{\bibfnamefont{A.~A.} \bibnamefont{{Abdo}}},
  \bibinfo{author}{\bibfnamefont{M.}~\bibnamefont{{Ajello}}},
  \bibinfo{author}{\bibfnamefont{A.}~\bibnamefont{{Allafort}}},
  \bibinfo{author}{\bibfnamefont{L.}~\bibnamefont{{Baldini}}},
  \bibinfo{author}{\bibfnamefont{J.}~\bibnamefont{{Ballet}}},
  \bibinfo{author}{\bibfnamefont{G.}~\bibnamefont{{Barbiellini}}},
  \bibinfo{author}{\bibfnamefont{M.~G.} \bibnamefont{{Baring}}},
  \bibinfo{author}{\bibfnamefont{D.}~\bibnamefont{{Bastieri}}},
  \bibinfo{author}{\bibfnamefont{A.}~\bibnamefont{{Belfiore}}},
  \bibinfo{author}{\bibfnamefont{R.}~\bibnamefont{{Bellazzini}}},
  \bibnamefont{et~al.}, \bibinfo{journal}{\apjs}
  \textbf{\bibinfo{volume}{208}}, \bibinfo{eid}{17} (\bibinfo{year}{2013}),
  \eprint{1305.4385}.

\bibitem[{\citenamefont{{Abdollahi} et~al.}(2022)\citenamefont{{Abdollahi},
  {Acero}, {Baldini}, {Ballet}, {Bastieri}, {Bellazzini}, {Berenji},
  {Berretta}, {Bissaldi}, {Blandford} et~al.}}]{Fermi12yr}
\bibinfo{author}{\bibfnamefont{S.}~\bibnamefont{{Abdollahi}}},
  \bibinfo{author}{\bibfnamefont{F.}~\bibnamefont{{Acero}}},
  \bibinfo{author}{\bibfnamefont{L.}~\bibnamefont{{Baldini}}},
  \bibinfo{author}{\bibfnamefont{J.}~\bibnamefont{{Ballet}}},
  \bibinfo{author}{\bibfnamefont{D.}~\bibnamefont{{Bastieri}}},
  \bibinfo{author}{\bibfnamefont{R.}~\bibnamefont{{Bellazzini}}},
  \bibinfo{author}{\bibfnamefont{B.}~\bibnamefont{{Berenji}}},
  \bibinfo{author}{\bibfnamefont{A.}~\bibnamefont{{Berretta}}},
  \bibinfo{author}{\bibfnamefont{E.}~\bibnamefont{{Bissaldi}}},
  \bibinfo{author}{\bibfnamefont{R.~D.} \bibnamefont{{Blandford}}},
  \bibnamefont{et~al.}, \bibinfo{journal}{\apjs}
  \textbf{\bibinfo{volume}{260}}, \bibinfo{eid}{53} (\bibinfo{year}{2022}),
  \eprint{2201.11184}.

\bibitem[{\citenamefont{{Ardavan}}(2023)}]{Ardavan23}
\bibinfo{author}{\bibfnamefont{H.}~\bibnamefont{{Ardavan}}},
  \bibinfo{journal}{Journal of High Energy Astrophysics}
  \textbf{\bibinfo{volume}{37}}, \bibinfo{pages}{62} (\bibinfo{year}{2023}),
  \eprint{2212.01305}.

\bibitem[{\citenamefont{{Olausen} and {Kaspi}}(2014)}]{Kaspi14}
\bibinfo{author}{\bibfnamefont{S.~A.} \bibnamefont{{Olausen}}}
  \bibnamefont{and} \bibinfo{author}{\bibfnamefont{V.~M.}
  \bibnamefont{{Kaspi}}}, \bibinfo{journal}{\apjs}
  \textbf{\bibinfo{volume}{212}}, \bibinfo{eid}{6} (\bibinfo{year}{2014}),
  \eprint{1309.4167}.

\bibitem[{\citenamefont{{Ardavan}}(2021)}]{Ardavan21}
\bibinfo{author}{\bibfnamefont{H.}~\bibnamefont{{Ardavan}}},
  \bibinfo{journal}{\mnras} \textbf{\bibinfo{volume}{507}},
  \bibinfo{pages}{4530} (\bibinfo{year}{2021}), \eprint{2104.06126}.

\bibitem[{\citenamefont{{Melrose} et~al.}(2021)\citenamefont{{Melrose},
  {Rafat}, and {Mastrano}}}]{Melrose}
\bibinfo{author}{\bibfnamefont{D.~B.} \bibnamefont{{Melrose}}},
  \bibinfo{author}{\bibfnamefont{M.~Z.} \bibnamefont{{Rafat}}},
  \bibnamefont{and}
  \bibinfo{author}{\bibfnamefont{A.}~\bibnamefont{{Mastrano}}},
  \bibinfo{journal}{\mnras} \textbf{\bibinfo{volume}{500}},
  \bibinfo{pages}{4530} (\bibinfo{year}{2021}), \eprint{2006.15243}.

\bibitem[{\citenamefont{{Ardavan} et~al.}(2008)\citenamefont{{Ardavan},
  {Ardavan}, {Singleton}, and {Perez}}}]{Ardavan09}
\bibinfo{author}{\bibfnamefont{H.}~\bibnamefont{{Ardavan}}},
  \bibinfo{author}{\bibfnamefont{A.}~\bibnamefont{{Ardavan}}},
  \bibinfo{author}{\bibfnamefont{J.}~\bibnamefont{{Singleton}}},
  \bibnamefont{and} \bibinfo{author}{\bibfnamefont{M.~R.}
  \bibnamefont{{Perez}}}, \bibinfo{journal}{\mnras}
  \textbf{\bibinfo{volume}{388}}, \bibinfo{pages}{873} (\bibinfo{year}{2008}),
  \eprint{0804.3123}.

\bibitem[{\citenamefont{{Singleton} et~al.}(2009)\citenamefont{{Singleton},
  {Sengupta}, {Middleditch}, {Graves}, {Perez}, {Ardavan}, and
  {Ardavan}}}]{Singleton}
\bibinfo{author}{\bibfnamefont{J.}~\bibnamefont{{Singleton}}},
  \bibinfo{author}{\bibfnamefont{P.}~\bibnamefont{{Sengupta}}},
  \bibinfo{author}{\bibfnamefont{J.}~\bibnamefont{{Middleditch}}},
  \bibinfo{author}{\bibfnamefont{T.~L.} \bibnamefont{{Graves}}},
  \bibinfo{author}{\bibfnamefont{M.~R.} \bibnamefont{{Perez}}},
  \bibinfo{author}{\bibfnamefont{H.}~\bibnamefont{{Ardavan}}},
  \bibnamefont{and}
  \bibinfo{author}{\bibfnamefont{A.}~\bibnamefont{{Ardavan}}},
  \bibinfo{journal}{arXiv e-prints} \bibinfo{eid}{arXiv:0912.0350}
  (\bibinfo{year}{2009}), \eprint{0912.0350}.

\bibitem[{\citenamefont{{Middleditch}}(2019)}]{Middleditch}
\bibinfo{author}{\bibfnamefont{J.}~\bibnamefont{{Middleditch}}},
  \bibinfo{journal}{arXiv e-prints} \bibinfo{eid}{arXiv:1910.03789}
  (\bibinfo{year}{2019}), \eprint{1910.03789}.

\bibitem[{\citenamefont{{Efstathiou} et~al.}(1988)\citenamefont{{Efstathiou},
  {Ellis}, and {Peterson}}}]{SWML}
\bibinfo{author}{\bibfnamefont{G.}~\bibnamefont{{Efstathiou}}},
  \bibinfo{author}{\bibfnamefont{R.~S.} \bibnamefont{{Ellis}}},
  \bibnamefont{and} \bibinfo{author}{\bibfnamefont{B.~A.}
  \bibnamefont{{Peterson}}}, \bibinfo{journal}{\mnras}
  \textbf{\bibinfo{volume}{232}}, \bibinfo{pages}{431} (\bibinfo{year}{1988}).

\bibitem[{\citenamefont{{Desai}}(2016)}]{Desai16}
\bibinfo{author}{\bibfnamefont{S.}~\bibnamefont{{Desai}}},
  \bibinfo{journal}{\apss} \textbf{\bibinfo{volume}{361}}, \bibinfo{eid}{138}
  (\bibinfo{year}{2016}), \eprint{1512.05962}.

\bibitem[{\citenamefont{{Lee} and {Petrosian}}(1997)}]{LeePetrosian}
\bibinfo{author}{\bibfnamefont{T.~T.} \bibnamefont{{Lee}}} \bibnamefont{and}
  \bibinfo{author}{\bibfnamefont{V.}~\bibnamefont{{Petrosian}}},
  \bibinfo{journal}{\apj} \textbf{\bibinfo{volume}{474}}, \bibinfo{pages}{37}
  (\bibinfo{year}{1997}), \eprint{astro-ph/9607127}.

\bibitem[{\citenamefont{{Maloney} and {Petrosian}}(1999)}]{Maloney}
\bibinfo{author}{\bibfnamefont{A.}~\bibnamefont{{Maloney}}} \bibnamefont{and}
  \bibinfo{author}{\bibfnamefont{V.}~\bibnamefont{{Petrosian}}},
  \bibinfo{journal}{\apj} \textbf{\bibinfo{volume}{518}}, \bibinfo{pages}{32}
  (\bibinfo{year}{1999}), \eprint{astro-ph/9807166}.

\bibitem[{\citenamefont{{Wheatland}}(2000)}]{Wheatland}
\bibinfo{author}{\bibfnamefont{M.~S.} \bibnamefont{{Wheatland}}},
  \bibinfo{journal}{\solphys} \textbf{\bibinfo{volume}{191}},
  \bibinfo{pages}{381} (\bibinfo{year}{2000}).

\bibitem[{\citenamefont{{Kocevski} and {Liang}}(2006)}]{Kocevski}
\bibinfo{author}{\bibfnamefont{D.}~\bibnamefont{{Kocevski}}} \bibnamefont{and}
  \bibinfo{author}{\bibfnamefont{E.}~\bibnamefont{{Liang}}},
  \bibinfo{journal}{\apj} \textbf{\bibinfo{volume}{642}}, \bibinfo{pages}{371}
  (\bibinfo{year}{2006}), \eprint{astro-ph/0601146}.

\bibitem[{\citenamefont{{Dainotti} et~al.}(2013)\citenamefont{{Dainotti},
  {Petrosian}, {Singal}, and {Ostrowski}}}]{Dainotti13}
\bibinfo{author}{\bibfnamefont{M.~G.} \bibnamefont{{Dainotti}}},
  \bibinfo{author}{\bibfnamefont{V.}~\bibnamefont{{Petrosian}}},
  \bibinfo{author}{\bibfnamefont{J.}~\bibnamefont{{Singal}}}, \bibnamefont{and}
  \bibinfo{author}{\bibfnamefont{M.}~\bibnamefont{{Ostrowski}}},
  \bibinfo{journal}{\apj} \textbf{\bibinfo{volume}{774}}, \bibinfo{eid}{157}
  (\bibinfo{year}{2013}), \eprint{1307.7297}.

\bibitem[{\citenamefont{{Dainotti} et~al.}(2015)\citenamefont{{Dainotti},
  {Petrosian}, {Willingale}, {O'Brien}, {Ostrowski}, and
  {Nagataki}}}]{Dainotti15}
\bibinfo{author}{\bibfnamefont{M.}~\bibnamefont{{Dainotti}}},
  \bibinfo{author}{\bibfnamefont{V.}~\bibnamefont{{Petrosian}}},
  \bibinfo{author}{\bibfnamefont{R.}~\bibnamefont{{Willingale}}},
  \bibinfo{author}{\bibfnamefont{P.}~\bibnamefont{{O'Brien}}},
  \bibinfo{author}{\bibfnamefont{M.}~\bibnamefont{{Ostrowski}}},
  \bibnamefont{and}
  \bibinfo{author}{\bibfnamefont{S.}~\bibnamefont{{Nagataki}}},
  \bibinfo{journal}{\mnras} \textbf{\bibinfo{volume}{451}},
  \bibinfo{pages}{3898} (\bibinfo{year}{2015}), \eprint{1506.00702}.

\bibitem[{\citenamefont{{Dainotti} et~al.}(2022)\citenamefont{{Dainotti},
  {Bargiacchi}, {Lenart}, {Capozziello}, {{\'O} Colg{\'a}in}, {Solomon},
  {Stojkovic}, and {Sheikh-Jabbari}}}]{Dainotti22}
\bibinfo{author}{\bibfnamefont{M.~G.} \bibnamefont{{Dainotti}}},
  \bibinfo{author}{\bibfnamefont{G.}~\bibnamefont{{Bargiacchi}}},
  \bibinfo{author}{\bibfnamefont{A.~{\L}.} \bibnamefont{{Lenart}}},
  \bibinfo{author}{\bibfnamefont{S.}~\bibnamefont{{Capozziello}}},
  \bibinfo{author}{\bibfnamefont{E.}~\bibnamefont{{{\'O} Colg{\'a}in}}},
  \bibinfo{author}{\bibfnamefont{R.}~\bibnamefont{{Solomon}}},
  \bibinfo{author}{\bibfnamefont{D.}~\bibnamefont{{Stojkovic}}},
  \bibnamefont{and} \bibinfo{author}{\bibfnamefont{M.~M.}
  \bibnamefont{{Sheikh-Jabbari}}}, \bibinfo{journal}{\apj}
  \textbf{\bibinfo{volume}{931}}, \bibinfo{eid}{106} (\bibinfo{year}{2022}),
  \eprint{2203.12914}.

\bibitem[{\citenamefont{{Bargiacchi} et~al.}(2023)\citenamefont{{Bargiacchi},
  {Dainotti}, and {Capozziello}}}]{Bargiacchi23}
\bibinfo{author}{\bibfnamefont{G.}~\bibnamefont{{Bargiacchi}}},
  \bibinfo{author}{\bibfnamefont{M.~G.} \bibnamefont{{Dainotti}}},
  \bibnamefont{and}
  \bibinfo{author}{\bibfnamefont{S.}~\bibnamefont{{Capozziello}}},
  \bibinfo{journal}{\mnras} \textbf{\bibinfo{volume}{525}},
  \bibinfo{pages}{3104} (\bibinfo{year}{2023}), \eprint{2307.15359}.

\bibitem[{\citenamefont{{Bagchi}}(2013)}]{Bagchi}
\bibinfo{author}{\bibfnamefont{M.}~\bibnamefont{{Bagchi}}},
  \bibinfo{journal}{International Journal of Modern Physics D}
  \textbf{\bibinfo{volume}{22}}, \bibinfo{eid}{1330021} (\bibinfo{year}{2013}),
  \eprint{1306.2152}.

\bibitem[{\citenamefont{{Bryant} et~al.}(2021)\citenamefont{{Bryant},
  {Osborne}, and {Shahmoradi}}}]{Bryant}
\bibinfo{author}{\bibfnamefont{C.~M.} \bibnamefont{{Bryant}}},
  \bibinfo{author}{\bibfnamefont{J.~A.} \bibnamefont{{Osborne}}},
  \bibnamefont{and}
  \bibinfo{author}{\bibfnamefont{A.}~\bibnamefont{{Shahmoradi}}},
  \bibinfo{journal}{\mnras} \textbf{\bibinfo{volume}{504}},
  \bibinfo{pages}{4192} (\bibinfo{year}{2021}), \eprint{2010.02935}.

\bibitem[{\citenamefont{{Dainotti} et~al.}(2021)\citenamefont{{Dainotti},
  {Petrosian}, and {Bowden}}}]{DPB21}
\bibinfo{author}{\bibfnamefont{M.~G.} \bibnamefont{{Dainotti}}},
  \bibinfo{author}{\bibfnamefont{V.}~\bibnamefont{{Petrosian}}},
  \bibnamefont{and} \bibinfo{author}{\bibfnamefont{L.}~\bibnamefont{{Bowden}}},
  \bibinfo{journal}{\apjl} \textbf{\bibinfo{volume}{914}}, \bibinfo{eid}{L40}
  (\bibinfo{year}{2021}), \eprint{2104.13555}.

\bibitem[{\citenamefont{{Manchester} et~al.}(2005)\citenamefont{{Manchester},
  {Hobbs}, {Teoh}, and {Hobbs}}}]{ATNF}
\bibinfo{author}{\bibfnamefont{R.~N.} \bibnamefont{{Manchester}}},
  \bibinfo{author}{\bibfnamefont{G.~B.} \bibnamefont{{Hobbs}}},
  \bibinfo{author}{\bibfnamefont{A.}~\bibnamefont{{Teoh}}}, \bibnamefont{and}
  \bibinfo{author}{\bibfnamefont{M.}~\bibnamefont{{Hobbs}}},
  \bibinfo{journal}{\aj} \textbf{\bibinfo{volume}{129}}, \bibinfo{pages}{1993}
  (\bibinfo{year}{2005}), \eprint{astro-ph/0412641}.

\bibitem[{\citenamefont{{Manchester} et~al.}(2001)\citenamefont{{Manchester},
  {Lyne}, {Camilo}, {Bell}, {Kaspi}, {D'Amico}, {McKay}, {Crawford}, {Stairs},
  {Possenti} et~al.}}]{PMBS}
\bibinfo{author}{\bibfnamefont{R.~N.} \bibnamefont{{Manchester}}},
  \bibinfo{author}{\bibfnamefont{A.~G.} \bibnamefont{{Lyne}}},
  \bibinfo{author}{\bibfnamefont{F.}~\bibnamefont{{Camilo}}},
  \bibinfo{author}{\bibfnamefont{J.~F.} \bibnamefont{{Bell}}},
  \bibinfo{author}{\bibfnamefont{V.~M.} \bibnamefont{{Kaspi}}},
  \bibinfo{author}{\bibfnamefont{N.}~\bibnamefont{{D'Amico}}},
  \bibinfo{author}{\bibfnamefont{N.~P.~F.} \bibnamefont{{McKay}}},
  \bibinfo{author}{\bibfnamefont{F.}~\bibnamefont{{Crawford}}},
  \bibinfo{author}{\bibfnamefont{I.~H.} \bibnamefont{{Stairs}}},
  \bibinfo{author}{\bibfnamefont{A.}~\bibnamefont{{Possenti}}},
  \bibnamefont{et~al.}, \bibinfo{journal}{\mnras}
  \textbf{\bibinfo{volume}{328}}, \bibinfo{pages}{17} (\bibinfo{year}{2001}),
  \eprint{astro-ph/0106522}.

\bibitem[{\citenamefont{{Yao} et~al.}(2017)\citenamefont{{Yao}, {Manchester},
  and {Wang}}}]{YMW16}
\bibinfo{author}{\bibfnamefont{J.~M.} \bibnamefont{{Yao}}},
  \bibinfo{author}{\bibfnamefont{R.~N.} \bibnamefont{{Manchester}}},
  \bibnamefont{and} \bibinfo{author}{\bibfnamefont{N.}~\bibnamefont{{Wang}}},
  \bibinfo{journal}{\apj} \textbf{\bibinfo{volume}{835}}, \bibinfo{eid}{29}
  (\bibinfo{year}{2017}), \eprint{1610.09448}.

\bibitem[{\citenamefont{{Ivezi{\'c}} et~al.}(2014)\citenamefont{{Ivezi{\'c}},
  {Connolly}, {Vanderplas}, and {Gray}}}]{astroml}
\bibinfo{author}{\bibfnamefont{{\v Z}.}~\bibnamefont{{Ivezi{\'c}}}},
  \bibinfo{author}{\bibfnamefont{A.}~\bibnamefont{{Connolly}}},
  \bibinfo{author}{\bibfnamefont{J.}~\bibnamefont{{Vanderplas}}},
  \bibnamefont{and} \bibinfo{author}{\bibfnamefont{A.}~\bibnamefont{{Gray}}},
  \emph{\bibinfo{title}{Statistics, Data Mining and Machine Learning in
  Astronomy}} (\bibinfo{publisher}{Princeton University Press},
  \bibinfo{year}{2014}).

\end{thebibliography}
\end{document}